\documentclass{article} 
\usepackage[preprint]{spconf}
\copyrightnotice{\copyright\ IEEE 2019}
\toappear{To appear in {\it Proc.\ ICASSP 2019,
May 12-17, 2019, Brighton, UK}}

\usepackage{amsmath}
\DeclareMathAlphabet{\orgmathcal}{OMS}{cmsy}{m}{n}
\usepackage{newtxmath} 
\usepackage{bm}  
\usepackage{graphicx}
\graphicspath{{./}}
\usepackage{datetime}
\usepackage{caption}
\usepackage{subcaption}
\usepackage[percent]{overpic}
\usepackage{placeins} 
\usepackage{outlines}
\usepackage{enumitem}
\usepackage{tabularx,colortbl}
\usepackage{makecell}
\usepackage{paralist}
\usepackage{algorithmicx }
\usepackage{color}
\usepackage{cite}
\usepackage{cleveref}
\Crefname{figure}{Figure}{Figures}  
\crefname{figure}{Fig.}{Figs.}  

\newcommand{\NN}{N}

\begin{document}
\title{Super-Resolution DOA Estimation for Arbitrary Array Geometries \\using a Single Noisy Snapshot }

\name{Anupama Govinda Raj and James H. McClellan \thanks{Supported by the Franklin Foundation, John \& Marilu McCarty Chair.}}
%\thanks{{Copyright \copyright~2019 IEEE}}
\address{School of Electrical and Computer Engineering, 
	Georgia Institute of Technology,
	Atlanta, USA
	\\
	Email: agr6@gatech.edu, jim.mcclellan@ece.gatech.edu}
\maketitle

\begin{abstract}
\noindent
We address the problem of search-free
DOA estimation from a single noisy snapshot for sensor arrays of arbitrary geometry, by extending a method of gridless super-resolution beamforming to arbitrary arrays with noisy measurements. The primal atomic norm minimization problem is converted to a dual problem in which the periodic dual function is represented with a trigonometric polynomial using truncated Fourier series. The number of terms required for accurate  representation depends linearly on the distance of the farthest sensor from a reference. The dual problem is then expressed as a semidefinite program and solved in polynomial time. DOA estimates are obtained via polynomial rooting followed by a LASSO based approach to remove extraneous roots arising in root finding from noisy data, and then source amplitudes are recovered by least squares. Simulations using circular and random planar arrays show high resolution DOA estimation in white and colored noise scenarios.
\end{abstract}


\thispagestyle{empty}
\vspace*{-3mm}
%________________________________________________________________________
\section{Introduction}\label{sec:intro}

Direction-of-arrival (DOA) estimation can be very challenging when snapshots are limited and sources are coherent
as in the case of fast moving sources and multipath arrivals. 
Under these conditions, high resolution DOA methods such as MVDR and MUSIC \cite{capon_mvdr_1969,schmidt_music_1986} 
fail due to  inaccurate estimation of spatial covariance matrix and self signal cancellation.

Sparsity based methods for DOA estimation inspired by compressed sensing (CS) \cite{donoho_cs_2006,candes_cs_2006,gurbuz_cs_beamform_2008,xenaki_compressive_beamforming_2014 } 
can tackle coherent sources and single snapshot. 
However, the CS based approaches are limited by the finite discrete grid of angles used to form the basis, leading to the \textit{off-grid} problem 
\cite{chi_basis_mismatch_2011} when the source directions do not lie on the grid. 
To improve performance, greedy algorithms with a highly coherent dictionary (finer search grids) are used
in \cite{duarte_baraniuk_spectral_cs_2013,fannjiang_coherence_2012},
but they are computationally demanding.
The off-grid DOA approaches \cite{zhu_perturbedCS_2011,yang_robust_perturbedCS_2012,austin_dynamicdictionary_2013,hu_gridrefine_2012} applicable for arbitrary arrays use a Taylor series approximation of array steering vectors on fixed grids, or iterative methods with dynamic grids to tackle the grid mismatch. 
However, their performance and accuracy depends on the grid density or they require noncovex optimization.
Recent gridless \textit{super-resolution} approaches using convex optimization \cite{tang_cs_offgrid_2013,candes_math_theory_super_resoln_2014, candes_super_resoln_noisy_2013,xenaki_gridfree_2015} eliminate the off-grid problem by forming the basis in the continuous angle domain
and provide high accuracy, but they are not applicable to arbitrary geometries.

In this paper, we develop a search-free DOA estimation method for  \iffalse 2-D \fi arrays of arbitrary geometry under the challenging conditions of coherent sources and a single noisy snapshot. 
This extends our earlier work \cite{agr_jmc_spl_SRarbarrays} on super-resolution DOA estimation for arbitrary geometry, to noisy measurements.
The DOA estimation problem for arbitrary geometry is solved as a dual maximization problem.  
By exploiting the periodicity and band-limited nature of the dual function, we can represent it with a \emph{finite trigonometric polynomial} using Fourier series (FS).
The proposed approach is motivated by \cite{rubsamen_gershman_FDrootmusic_2009, doron_wavefield_1994,doron_wavefield_algos_1994,belloni_MS_arbarrays_2007},
where root-MUSIC is extended to arbitrary arrays.
The modified dual problem can then be expressed as a \emph{finite} semidefinite program (SDP), and solved.
Finally, the search-free DOA estimates are obtained through polynomial rooting of a nonnegative polynomial formed from the dual polynomial.
To remove the extraneous roots arising in the noisy case, 
we use a LASSO-like approach related to \cite{tan_nehorai_SRcoprime_2014,hung_kaveh_SRmultisnaps_2015}.
%________________________________________________________________________
\vspace*{-7mm}

\section{Data Model} 
Consider an $ M $-element array of arbitrary geometry, which receives signals from $ L $ narrowband far-field sources with complex amplitude $ s_l $ and azimuth DOA $ \theta_l $, 
$ l = 1, \dots, L $.
We define the sparse source function $ x(\theta) $ in the continuous angle domain $\theta\in(-\pi,\pi]$ with impulses 
as
$   x(\theta) =  \sum_{l=1}^{L} s_l \delta(\theta- \theta_l) $.
Then the $ M \times 1 $ observed array snapshot vector $\bm{y} $ is 
\vspace*{-2mm}
\begin{equation}
	\label{eq:MeasurementModel}
	\bm{y} = \mathcal{S} x + \bm{n}, \;\text{where}\; y_m = n_m + \int\limits_{-\pi}^{\pi} a_m(\theta) x(\theta) d\theta, 
\end{equation} 
$ m=1,\dots,M  $ and $\bm{n} \in \mathbb{C}^{M} $ is the received additive noise across the array. 
The linear measurement operator $ \mathcal{S} $ represents the array manifold over $\theta$, whose $ m $-th component $a_m(\theta)$   
is the response of the $ m $-th sensor for a source at direction $\theta$.
\begin{equation}
	a_m(\theta) = e^{-j2\pi f \tau_m(\theta)},
	\label{eq:ar-spDefined}
\end{equation}
where $ \tau_m(\theta)$ is the propagation delay with respect to a reference.%
\footnote{We prefer to study $a_m(\theta)$ as a function of $\theta$. On the other hand, at a specific angle $\theta_1$, $[a_m(\theta_1)]\in\mathbb{C}^{M}$ is the steering vector for direction $\theta_1$.}
For narrowband sources of frequency $f$ and propagation speed $v$, the wavelength is $\lambda = v/f$. 
Using $ \tau_m(\theta) = {\langle\bm{p}_m, \bm{u}_{\theta}\rangle}/{v}$, we simplify the exponent in \eqref{eq:ar-spDefined} as 
\begin{equation} 
    2\pi f \tau_m(\theta) 
  = 2\pi (|\bm{p}_m|/\lambda) \cos(\theta - \angle \bm{p}_m),
  \label{eq:ar-spExponent}
\end{equation} 
where $ \bm{p}_m $ is the position vector of the $ m $-th sensor with respect to a reference, and $\bm{u}_{\theta}$ is a unit vector in direction $\theta$.

\vspace{-3mm}

\section{Proposed Method}
Assuming the sources are sparse in angle, $x(\theta)$ 
could be recovered from noisy measurements $\bm{y} = \mathcal{S} x + \bm{n}$ via  \cite{candes_super_resoln_noisy_2013}
\begin{equation}\label{eq:gridfree_primal}
	\min_x \, \lVert x \rVert^{\phantom{.}}_{\orgmathcal{A}}, \quad {\rm s.t.} \quad \lVert\bm{y} - \mathcal{S} x  \rVert^{\phantom{.}}_2 \le \delta,
\end{equation}
where $ \delta $ satisfies the condition that  $ \lVert \bm{n} \rVert^{\phantom{.}}_2  \le \delta $. $ \lVert . \rVert^{\phantom{.}}_{\orgmathcal{A}} $ denotes the {atomic norm} \cite{chandrasekaran_atomicnorm_2012} which is a continuous analogue of the $ l_1 $ norm, i.e., \(
\lVert x \rVert^{\phantom{.}}_{\orgmathcal{A}} = \sum_{l=1}^{L} \lvert s_l \rvert.
\)
Here $ \mathcal{S}  $ does not represent Fourier measurements, unlike \cite{tang_cs_offgrid_2013,candes_math_theory_super_resoln_2014,candes_super_resoln_noisy_2013,xenaki_gridfree_2015}.
The primal problem \eqref{eq:gridfree_primal} is infinite dimensional and difficult to solve. Therefore, we work with 
the corresponding dual maximization problem
\begin{equation}\label{eq:gridfree_dual}
	\max_{\bm{c}\in \mathbb{C}^M } \operatorname{\Re}\{\bm{c}^H \bm{y}\} - \delta {\lVert \bm{c}\rVert}_{2},\quad \text{s.t.\ \;}  \lVert \mathcal{S}(\theta)^H \bm{c} \rVert _{\infty} \le 1,
\end{equation}
where $\bm{c}$ is the dual variable (see details in \cite{xenaki_gridfree_2015, candes_super_resoln_noisy_2013 }). 
The \textit{dual function} defined by
$ \mathcal{S}(\theta)^H \bm{c} $ has unit magnitude in the direction of actual sources, irrespective of geometry.
For a uniform linear array (ULA), $ \mathcal{S}(\theta)^H \bm{c} $ is, in fact, an \mbox{$(M-1)^\text{th}$} degree polynomial in $e^{j\theta}$, and \eqref{eq:gridfree_dual} is then solved using an SDP \cite{candes_math_theory_super_resoln_2014, candes_super_resoln_noisy_2013,xenaki_gridfree_2015}. 
The polynomial structure arises from the fact that sensor delays, $\tau_m(\theta)$ in 
(\ref{eq:ar-spDefined}), for a ULA are integer multiples of a constant.
For arbitrary arrays, $ \mathcal{S}(\theta)^H \bm{c} $ cannot be directly expressed as a polynomial, but we overcome this difficulty with a Fourier domain (FD) representation of the dual function that provides a polynomial form for the SDP.

\vspace*{-3mm}
\subsection{Fourier Domain Representation of the Dual Function}
We review the Fourier series representation of the dual function \cite{agr_jmc_spl_SRarbarrays} here for completeness. 
The function $ b(\theta) = \mathcal{S}(\theta)^H \bm{c}$ is periodic in $ \theta $ with period $ 2\pi $ as it is a linear combination of smooth (band-limited) periodic functions, $a^*_m(\theta)$, $m=1,\dots, M$. 
Thus, $ b(\theta) $ has a Fourier series (FS)
which can be truncated if its Fourier coefficients $ {B}_{k}\!\approx\!0 $ for $|k|>N$.
Each $a^*_m(\theta)$, being periodic, has a FS with coefficients $\alpha_m[k]$, related to $B_k$ via $B_k = \sum_m \alpha_m[k] c_m$.
So we have
\begin{equation}
  b(\theta) 
  = \sum_{k=-N}^{N} \sum_{m=1}^{M}(\alpha_m[k] c_m) e^{jk\theta},
  \label{eq:BkFromAlphak}
\end{equation}
which is a \emph{finite degree} polynomial in $ z = e^{j\theta} $.
As a result, we can determine $N$ for FS truncation by examining the FS coefficients at each sensor, $\alpha^{\phantom{.}}_m[k]$, which depend solely on the array geometry and not on the measured signals. 

Assuming a sufficiently large number of DFT points $(P=2N+1)$ for dense sampling in $\theta$, the FS coefficients $\alpha^{\phantom{.}}_m[k]$ can be estimated from samples of $a^*_m(\theta)$ using the DFT \cite{rubsamen_gershman_FDrootmusic_2009,DSP_First_2015} as, 
\begin{equation}
 \hat\alpha^{\phantom{.}}_m[k] \simeq
    (1/P)\sum_{l=-N}^{N} a^*_m(l\Delta\theta) e^{-j(2\pi/P) l k},
   \label{eq:P-ptDFT}
\end{equation}
where $\Delta\theta=2\pi/P$, and $k=-N,\ldots,0,1,\ldots,N$.
Note that circular indexing of the DFT is exploited in \eqref{eq:P-ptDFT}.

\begin{figure}[htbp]
	\begin{subfigure}{.5\columnwidth}
		\centering
		\begin{overpic}[width=1\textwidth]{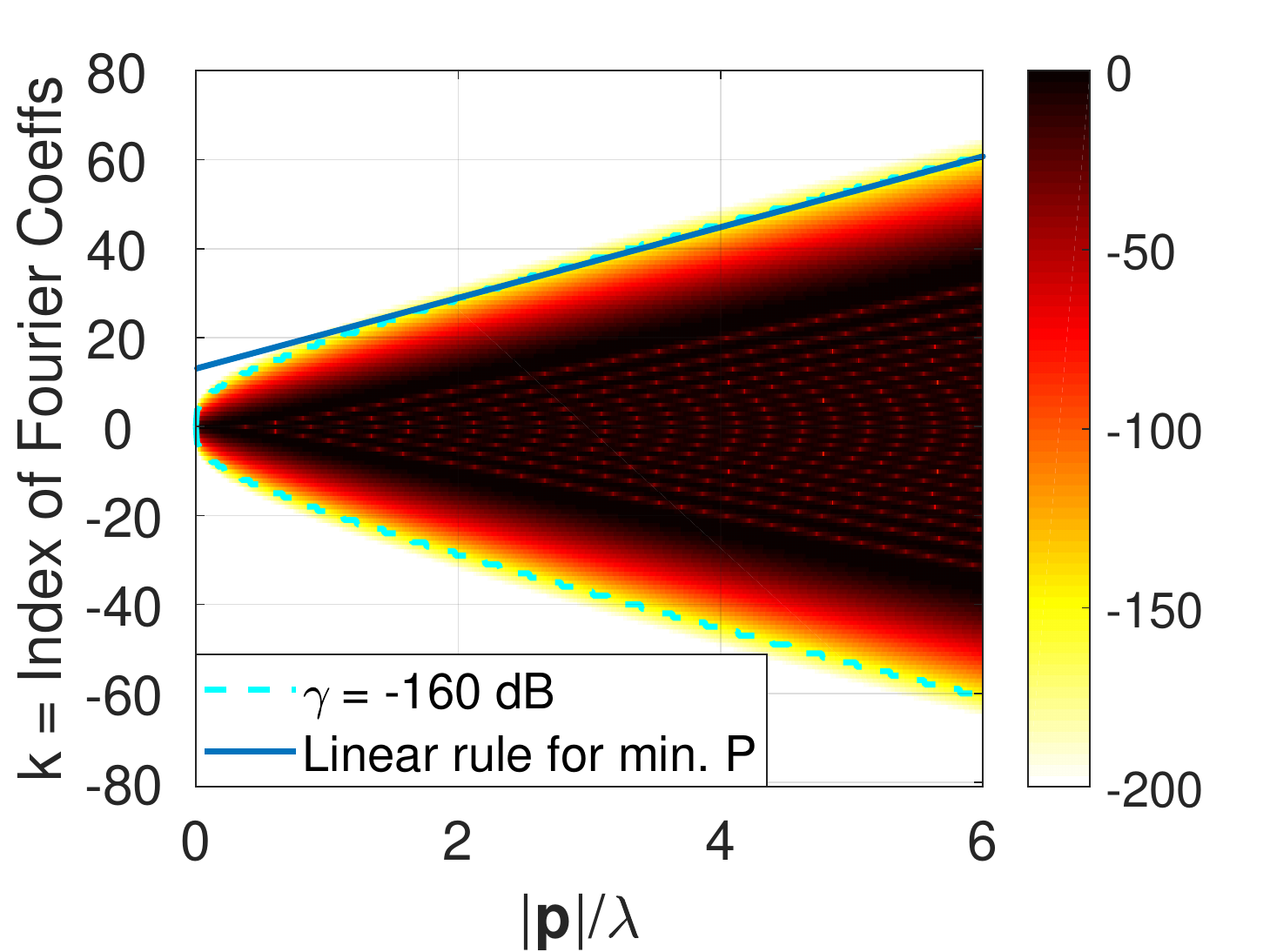}
			\put (17,63 ) {{\small {(a)}}}
		\end{overpic}
		\label{fig:Fig1a}
	\end{subfigure}%
	\begin{subfigure}{.5\columnwidth}
		\centering
		\begin{overpic}[width=1\textwidth]{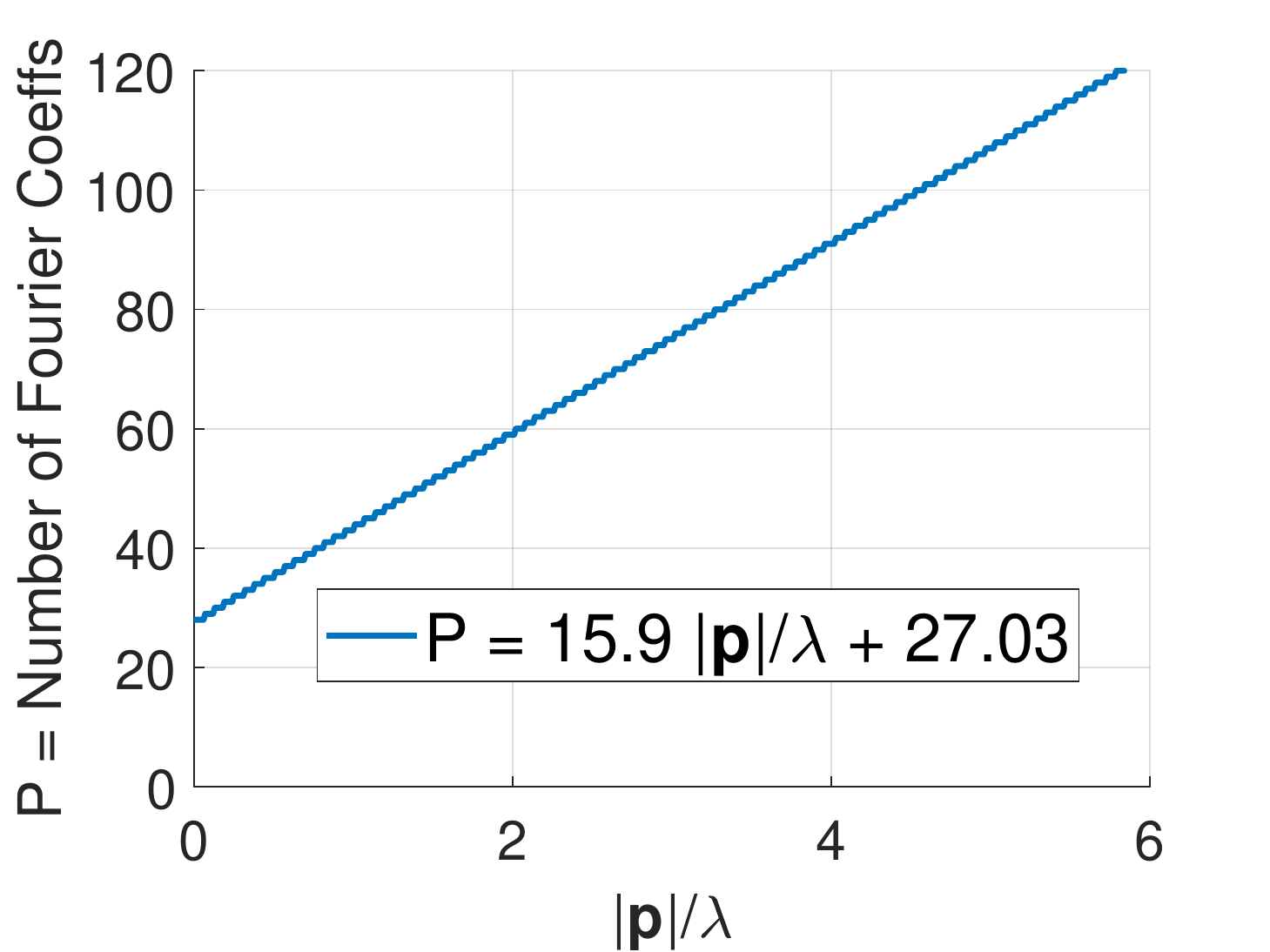}
			\put (17,63 ) {{\small {(b)}}}
		\end{overpic}
		\label{fig:Fig1b}
	\end{subfigure}
\vspace*{-6mm}
	\caption{ (a) Squared magnitude (dB) of FS coefficients as a function of $k$, the DFT index  and $ |\bm{p}|/\lambda $, the normalized distance of sensor from origin, (b) minimum $ P $ for good DFT approximation vs.~$ |\bm{p}|/\lambda $, for FS magnitude cutoff $\gamma = -160 $\,dB.}
	\label{fig:Fig1}
\end{figure}

Next, we conduct a numerical study of the FS for the continuous function $a^*_m(\theta)$ defined in (\ref{eq:ar-spDefined},\,\ref{eq:ar-spExponent}) to determine the value of $P$ needed for various array geometries.
The FS coefficients of  
$ a^*_m(\theta) $ can be approximated numerically by a very long DFT to get $ \hat\alpha[k] $.
From (\ref{eq:ar-spDefined},\,\ref{eq:ar-spExponent}), the magnitude $|\hat\alpha_m[k]|$ depends only on $ |\bm{p}|/\lambda $, the normalized distance of the sensor from the origin. 
This is because $(\theta-\angle\bm{p})$ is a shift in the argument of $a_m(\theta)$ which changes only the phase of its FS coefficients.
We use a long DFT to obtain FS coefficients for many different values of $ |\bm{p}|/\lambda $, and display the magnitude $|\hat\alpha[k]|^2$ as an image in  \cref{fig:Fig1}, which confirms that $\hat\alpha[k]$ is bandlimited.
\vspace{-2mm}
We observe that as  $ |\bm{p}| / \lambda $ increases, the bandwidth of the FS grows and hence the distance of the farthest sensor from origin in an array
controls the minimum $ P $ needed to get an accurate DFT representation.
The index $N$ where $|\hat\alpha[k]| \approx 0$ for $|k|>N$  depends on choosing a threshold $ \gamma$ for 
the squared magnitude of the FS. 
\Cref{fig:Fig1} shows the case for $\gamma = -160$\,dB below the maximum.
A linear approximation derived for $ |\bm{p}| / \lambda \geq 2$ gives an excellent estimate for $P = 2N+1$.
This minimum value of $ P $ is important for reducing the computational complexity of the SDP.
For $\gamma =-160$\,dB, the linear estimate is, 
\begin{equation}
P = 15.9 |\bm{p}| / \lambda +27.03.
\end{equation}
An example verifying the match between predicted and observed $ P $ needed to ensure success is presented in \cite{agr_jmc_spl_SRarbarrays}.

Using the DFT representation in \eqref{eq:P-ptDFT}, the dual function $ b(\theta) $ can be related to a \textit{dual polynomial} $ \hat b(z) $ as
\begin{align}\label{eq:dualfunction_approx}
	b(\theta) 
	&\simeq \sum_{k=-\NN}^{\NN}\hat{B}_{k}e^{jk\theta}
	= \sum_{k=-\NN}^{\NN}\hat{B}_{k}z^{k} 
	\;\buildrel{\Delta}\over{=}\;\hat b(z)\bigg{|}_{z=e^{j\theta}}
\end{align}
Combining \eqref{eq:BkFromAlphak} and \eqref{eq:dualfunction_approx}, we recognize that the coefficients $\hat{B}_{k}$ can be written in matrix-vector form with $\bm{h}\in\mathbb{C}^{P}$ being
\begin{equation}
	\label{eq:Gc}
	{\bm{h}} = \left[ \hat{B}_{-\NN} \;\;  \hat{B}_{-(\NN-1)} \;\ldots\;  \hat{B}_{\NN} \right]^T = \bm{G}^H \bm{c},
\end{equation}
where  $\bm{G}^H =  \begin{bmatrix} \hat\alpha_m[k] \end{bmatrix}_{P\times M}$ is a matrix  whose $m$-th column contains the FS coefficients of $a^*_m(\theta)$, and $\bm{c}$ is the dual vector.

%________________________________________________________________________
\vspace*{-2mm}

\subsection{Semidefinite Programming}
We convert the infinite number of constraints in the dual problem \eqref{eq:gridfree_dual} into finite-dimensional %semidefinite matrix inequalities 
matrix constraints as in \cite{xenaki_gridfree_2015,candes_super_resoln_noisy_2013}, by using the uniform boundedness of the function $\mathcal{S}(\theta)^H \bm{c}$ in \eqref{eq:gridfree_dual} and hence that of its FD representation given by the dual polynomial $\hat b(z) $, to obtain
the following SDP.
\begin{align*}\label{eq:FDgridfree_SDP}
\min_{\bm{c},\bm{H}} \operatorname{\Re}\{\bm{c}^H\bm{y}\}-&\delta {\lVert \bm{c}\rVert}_{2},\ 
	{\rm s.t.} 
	\begin{bmatrix} \bm{H}_{P \times P} &  \bm{G}^H_{P \times M}\bm{c}^{\phantom{.}}_{M \times 1} \\
	\bm{c}^H \bm{G} & 1 \end{bmatrix} \succeq 0,
	\tag{$ 11$}
	\\
	&\ \sum_{i=1}^{P-j}\bm{H}_{i,i+j} = \left\{ \begin{array}{cl}
	1, &\mbox{ $j = 0$} \\
	0 &\mbox{ $j = 1, \ldots , P-1$},
	\end{array}\right.
\end{align*}
where $ \operatorname{\Re}\{\cdot\} $ denotes the real part.
$ \bm{H}$ is a positive semidefinite matrix satisfying the constraints in \eqref{eq:FDgridfree_SDP}.
The SDP \eqref{eq:FDgridfree_SDP} has $ n = P^2/2 + M$ optimization variables and is solvable in polynomial time by interior-point methods \cite{vandenberghe_boyd_sdp_1996}.
The observed time complexity was found to be much less than the worst case $ \mathcal{O}(n^3) $.
The dual polynomial $\hat b(z)$ is the desired output after the SDP, so its coefficient vector is constructed from the optimal $ \bm{c}_* $ via 
$ \bm{h}_* = \bm{G}^H \bm{c}_* $. 
\vspace*{-2.3mm}

\subsection{Recovery of Source DOAs and Amplitudes } 
For sufficiently large $ P $, the approximation of the dual function $ \mathcal{S}(\theta) ^H \bm{c}_*  $ by the dual polynomial $\hat b(e^{j\theta}) $ is highly accurate.
Based on the constraint \eqref{eq:gridfree_dual}, $|\hat b(e^{j\theta})| $ would be equal to one for true DOAs, and less than one elsewhere
 \cite{candes_math_theory_super_resoln_2014}.
To locate the angles $ \theta $ where the magnitude of the dual polynomial is one, we form a nonnegative polynomial 
$ p(z) = 1- |\hat b(z)|^2  $
from the dual polynomial coefficients $ \bm{h}_* $.
The coefficients of $|\hat b(z)|^2$, denoted by $ r_k $, are the autocorrelation coefficients of $ \bm{h}_* $, i.e.,  $ r_k  = \sum_j h_j h_{j-k}^*	$.
The angles of the zeros of $ p(z) $ on the unit circle include the DOAs of the sources.

Due to numerical issues of polynomial rooting at low SNRs, the SDP might provide extraneous unit-circle zeros that do not correspond to true sources.
Therefore, the DOAs are finally recovered by  
a sparsity-promoting $\ell_1$ problem.
\begin{equation*}\label{eq:l1_min}
\min_{\bm{x}} \, \lVert \bm{x} \rVert^{\phantom{.}}_{1}, \quad {\rm s.t.} \quad \lVert\bm{y} - \bm{A}_{\text{aug}} \bm{x}  \rVert^{\phantom{.}}_2 \le \epsilon,
	\tag{$ 12$}
\end{equation*}
where $ \bm{A}_{\text{aug}}  $ is a dictionary of steering vectors that has steering vectors for a discrete set of angles as its columns. 
This discrete set includes the angles of the unit-circle roots from the SDP, as well as additional angles drawn from a uniform distribution in $ (-180^\circ,180^\circ] $.
Then \eqref{eq:l1_min} is written as the following LASSO-like problem and solved using convex optimization.
\begin{equation*}\label{eq:l1_lasso}
\min_{\bm{x}} \, \tfrac{1}{2}\lVert\bm{y} - \bm{A}_{\text{aug}} \bm{x}  \rVert^{\phantom{.}}_2 + \beta\lVert x \rVert^{\phantom{.}}_{1}   ,
	\tag{$ 13$}
\end{equation*}
The support of the solution $ \bm{x}_* $ yields the DOAs of interest.

Once we estimate the DOAs, the amplitudes of the sources are recovered by least squares 
$ \hat{\bm{s}} =  \bm{A}(\hat{\bm{\theta}})^\dagger \bm{y} $
where $ ^{\dagger} $ denotes the pseudo-inverse. The columns of the matrix $ \bm{A}(\hat{\bm{\theta}}) $ are the steering vectors for the estimated DOAs $ \hat{\bm{\theta}} $.

\vspace{1mm}\noindent To summarize, the \emph{steps} in the proposed method are: \\ 
{\ninept
1. For the geometry, compute $ \bm{G}^H = \begin{bmatrix} \hat\alpha_m[k] \end{bmatrix} $ via \eqref{eq:P-ptDFT}.\\ 
2. Estimate the noise level, and set $\delta$.\\ 
3. Using $ \bm{G} $, $ \bm{y} $ and $\delta$, 
solve the SDP in \eqref{eq:FDgridfree_SDP} to find the optimal $
\bm{c}_* $.\\
4. Get the optimal dual polynomial coefficients via $ \bm{h}_*  = \bm{G}^H
\bm{c}_*  $.\\
5. Estimate DOAs $ \hat{\bm{\theta}} $ by finding angles of unit circle roots of $ p(z) $.\\
6. Eliminate extraneous zeros via the $\ell_1$ sparsity optimization \eqref{eq:l1_lasso}.\\
7. Recover the source amplitudes $ \hat{\bm{s}} $ by least squares.
}

\section{Simulations}
Results for the uniform circular array (UCA) and random planar array (RPA) geometries 
are presented in \Cref{subsec:UCA,subsec:RPA}. 
Performance is compared with the conventional delay-sum beamformer (CBF). 
All simulations consider 
a single snapshot and multiple coherent sources \cite{shan_spatialsmoothing_1985}, which are complex sinusoids of the same frequency with constant phase difference.
We implemented the SDP \eqref{eq:FDgridfree_SDP} using CVX \cite{grant_boyd_cvx_2008}.
For DOA estimation, we use only those roots of $ p(z) $ that lie within a distance of $ 0.02 $ from the unit circle.
\vspace{-3mm}

\subsection{Simulations for Uniform Circular Array (UCA)}\label{subsec:UCA}
Two examples using a 40-element UCA are presented here. 
The array radius is $r= 2\lambda $, and the uniform sensor separation is $ d = (\pi/10)\lambda $.
With the reference point at the center of the array, $|\bm{p}_m| =r$ for all sensors.

\begin{figure}[htbp]
	\begin{subfigure}{.5\columnwidth}
		\centering
		\includegraphics[width=1\linewidth]{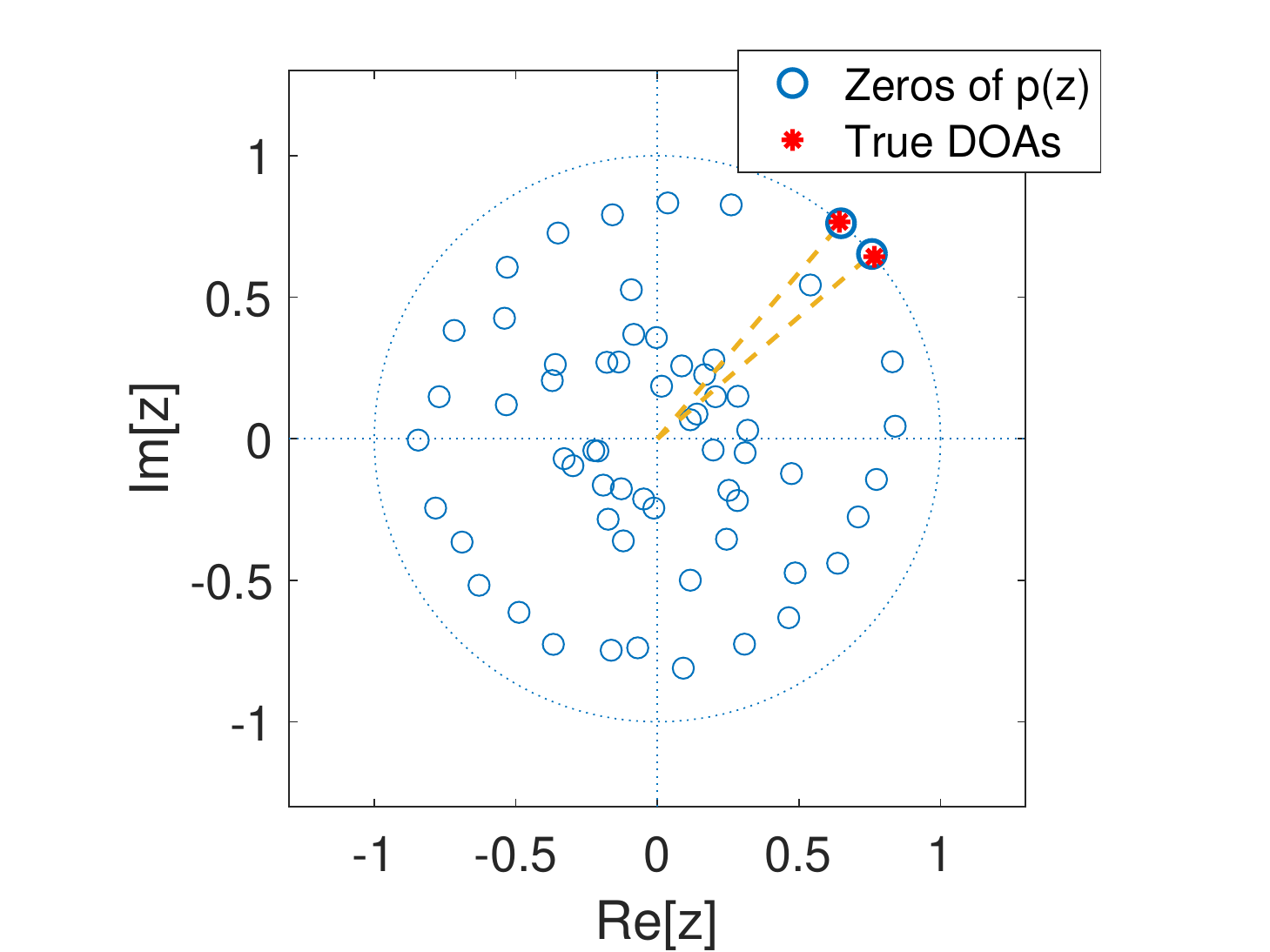}
		\caption{Zeros of $ p(z) $}
		\label{fig:Fig2a}
	\end{subfigure}%
	\begin{subfigure}{.5\columnwidth}
		\centering
		\includegraphics[width=1\linewidth]{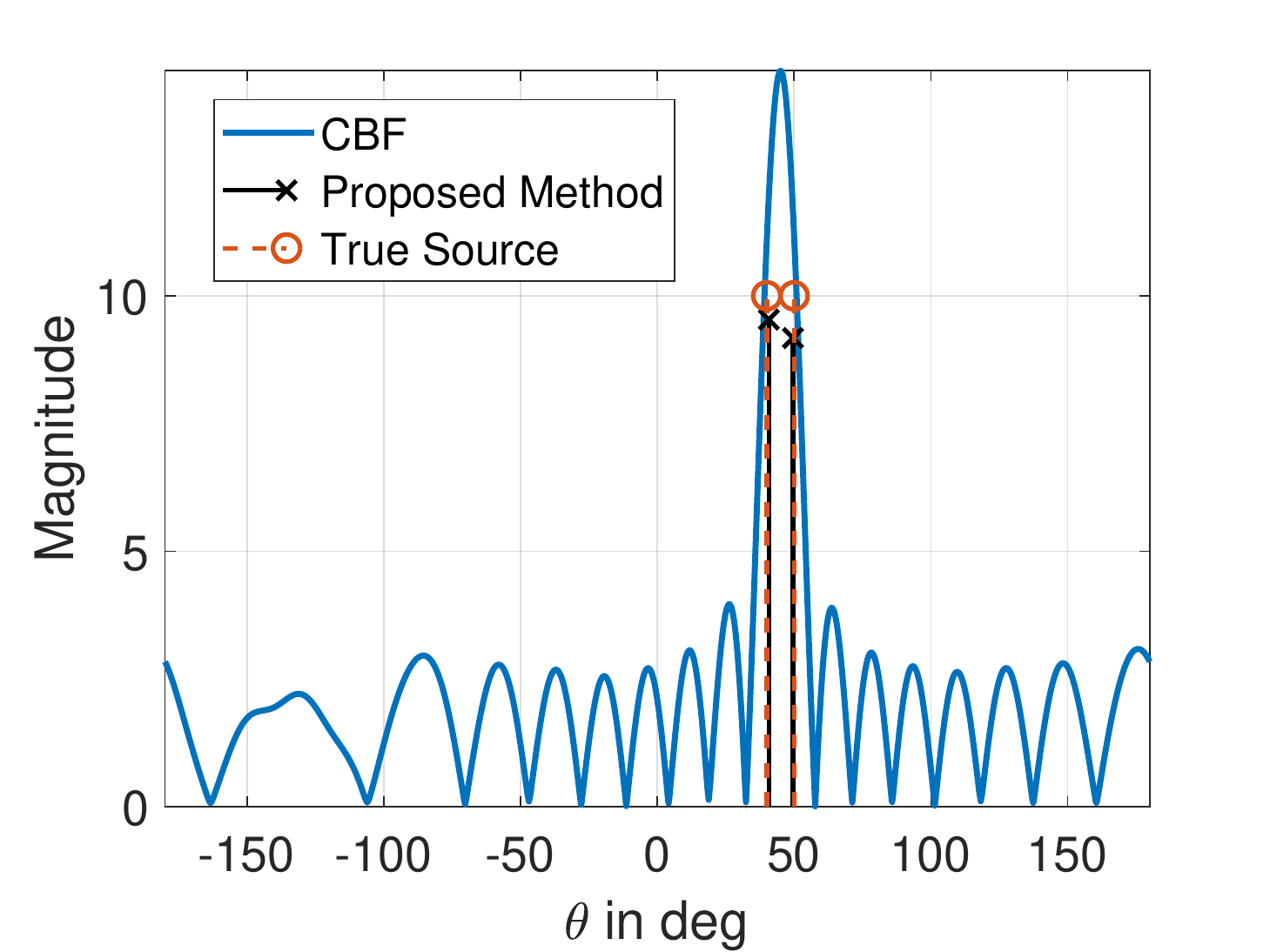}
		\caption{CBF vs. Proposed Method}
		\label{fig:Fig2b}
	\end{subfigure}
	\caption{Colored noise example : UCA with $M=40$, $P=63 $. Two sources at $ 40^\circ, 50^\circ$ with $ 20 $\,dB SNR.
	}
	\label{fig:Fig2}
\end{figure}

In the first example in \cref{fig:Fig2b}, we study the angular resolution of the proposed method by considering two equal magnitude sources of SNR $ 20 $\,dB separated by $ 10^\circ $.
Since the noise in practice is often colored, we simulate noise with $ 1/f $ spectral decay along frequency for this example.  
As seen in \cref{fig:Fig2b}, the CBF is not able to resolve the two closely located sources, whereas estimates from the unit-circle zeros in \cref{fig:Fig2a} are very accurate. 
This reinforces \iffalse the fact \fi that the proposed approach offers higher resolution than existing methods for single snapshot DOA estimation.
The approach only assumes additive noise and this example also verifies its applicability to colored noise scenarios.
Additive white Gaussian noise is used in the rest of the examples.

\begin{figure}[htbp]
	\begin{subfigure}{.48\columnwidth}
		\centering
		\includegraphics[width=1\linewidth]{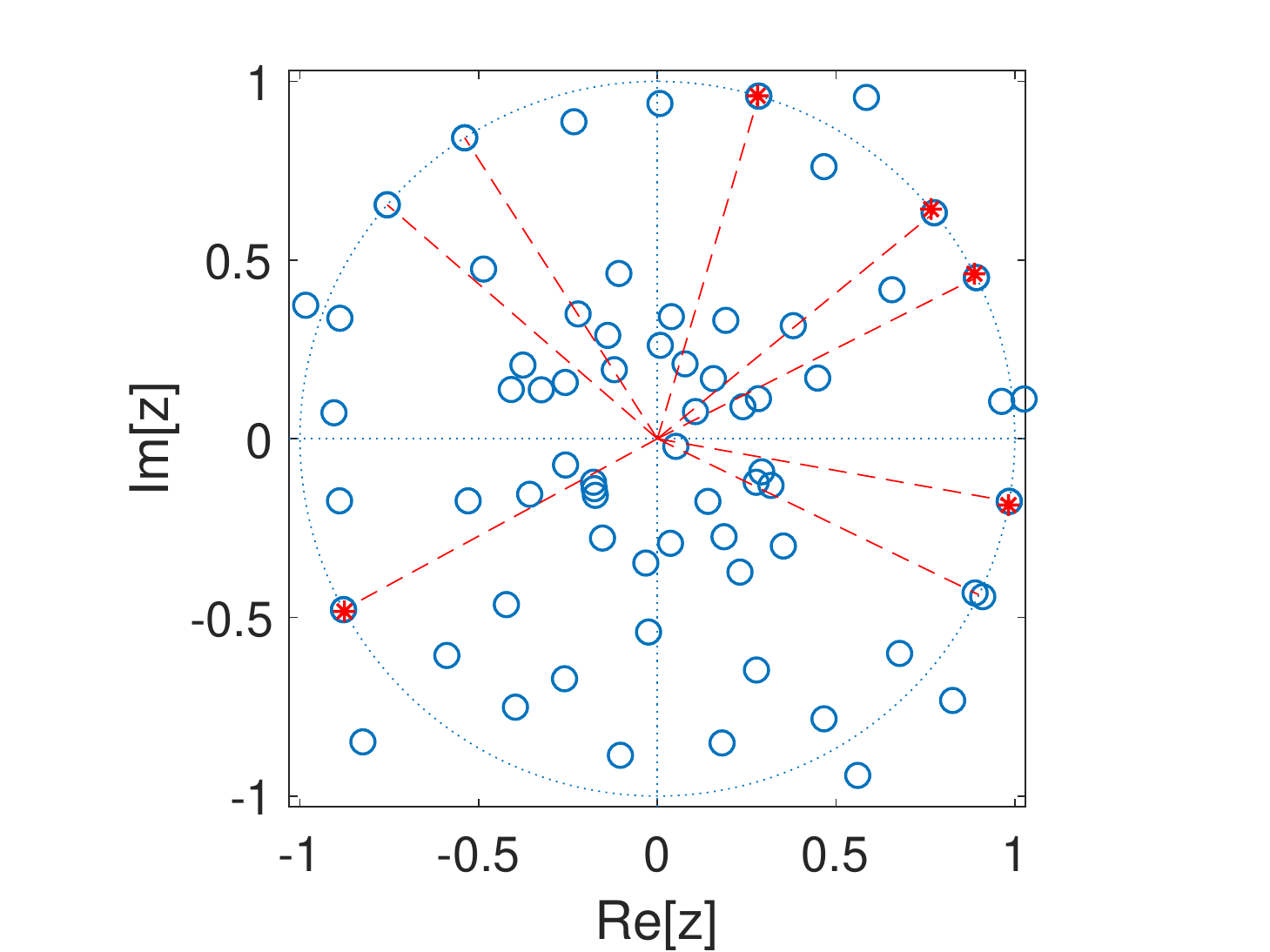}
		\caption{Zeros of $ p(z) $}
		\label{fig:Fig3a}
	\end{subfigure}
	\begin{subfigure}{.5\columnwidth}
		\centering
		\includegraphics[width=1\linewidth]{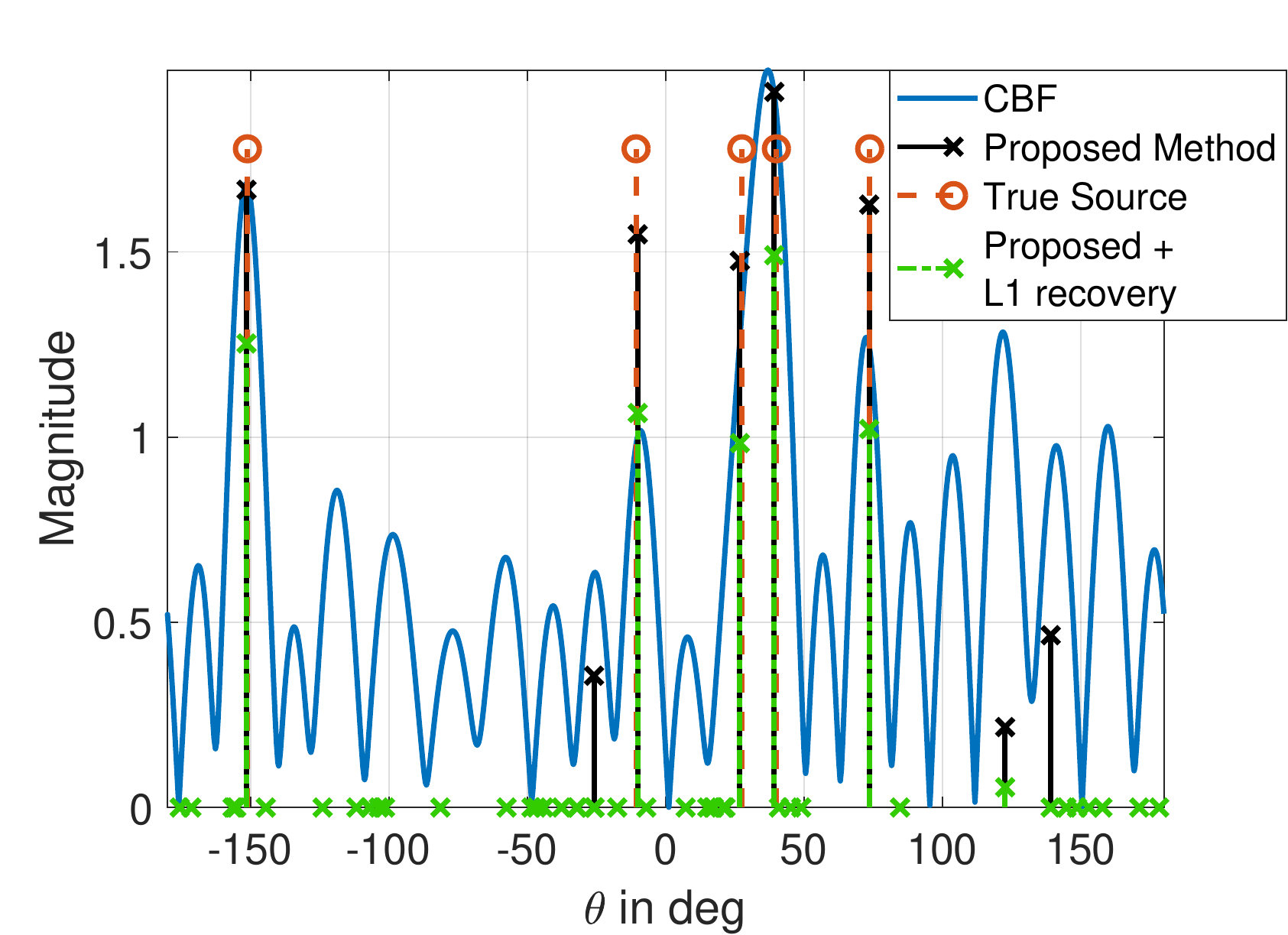}
		\caption{CBF vs. Proposed Method}
		\label{fig:Fig3b}
	\end{subfigure}
	\caption{Result for UCA with $M=40$, $P=63 $. Five sources at $ -10.7^\circ, 27.5^\circ, 40^\circ, 73.7^\circ$ and $-151.1^\circ$ of $ 5 $\,dB SNR. 
}
	\label{fig:Fig3}
\end{figure}

In \cref{fig:Fig3}, we consider five equal magnitude sources at $ 5 $\,dB SNR.
Due to the lower SNR, the set of unit-circle zeros of $p(z)$ in \cref{fig:Fig3a} includes three extraneous zeros in addition to the five zeros that correspond to the true DOAs.
Using the $ \ell_1 $ norm based DOA recovery in \eqref{eq:l1_lasso},
we eliminate those unwanted roots as shown in
\cref{fig:Fig3b}.
The nonzero elements in the  $ \ell_1 $ recovery result are the final estimated DOAs.
The amplitudes from the $ \ell_1 $ recovery are expected to be inaccurate due to shrinkage operation.
Once we estimate the DOAs, the amplitudes can be recovered via least squares.
The CBF is unable to resolve two among the five sources, and shows high side lobes as well inaccurate source amplitude estimates. 
On the other hand, the proposed approach in combination with the $ \ell_1 $ recovery accurately estimates the DOAs of all three sources. 
Both UCA examples validate the ability of the proposed method to estimate DOAs accurately for an arbitrary 2-D array. % with very high accuracy. %
\vspace*{-3mm}

\subsection{Simulation for Random Planar Array (RPA)}\label{subsec:RPA}
\begin{figure}[h!]
	\centering
	\begin{subfigure}{.49\columnwidth}
		\centering
		\includegraphics[width=1\linewidth]{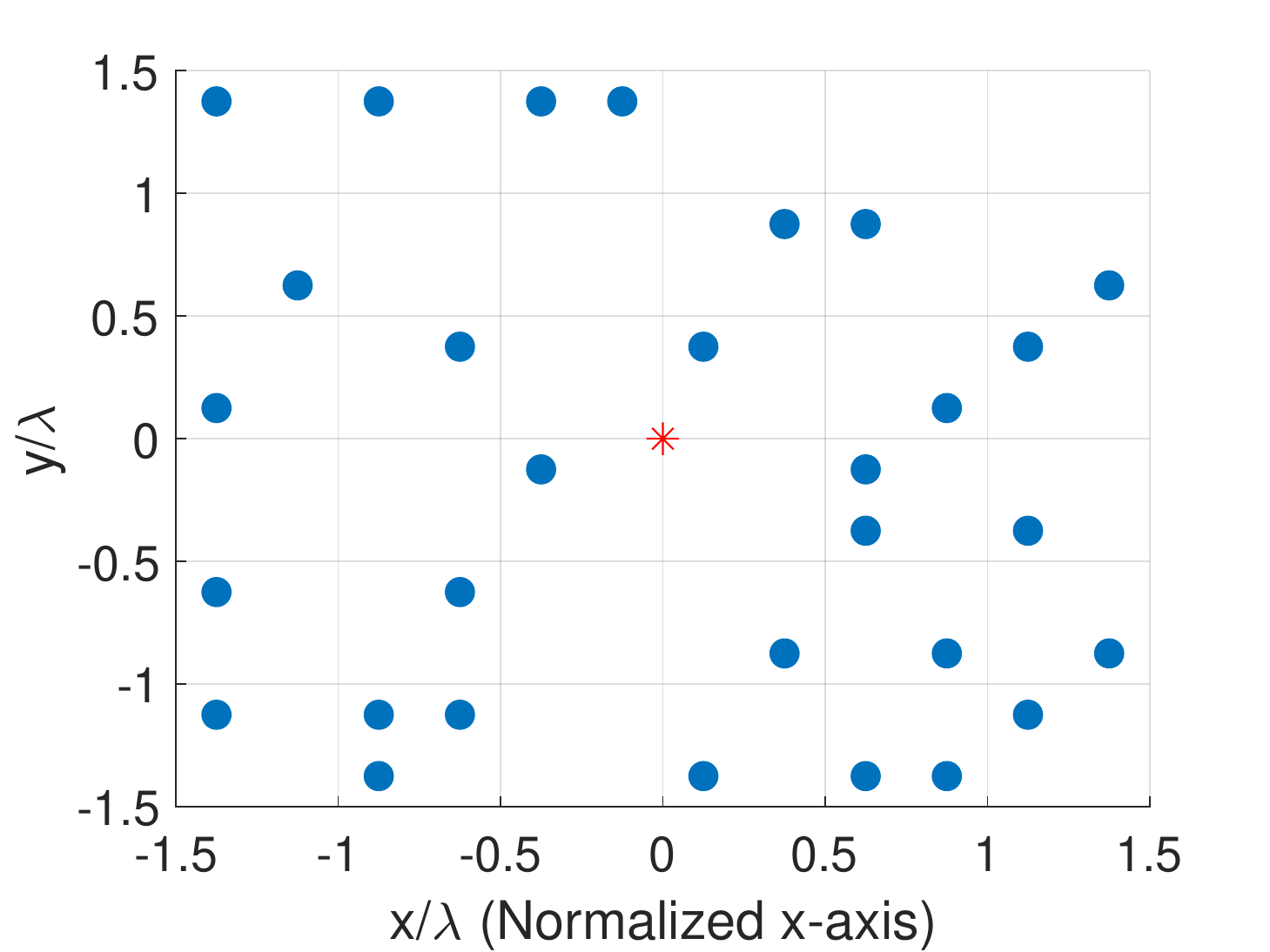}
		\caption{Random Planar Array (RPA)}
		\label{fig:Fig4a}
	\end{subfigure}
	\begin{subfigure}{.49\columnwidth}
		\centering
		\includegraphics[width=1\linewidth]{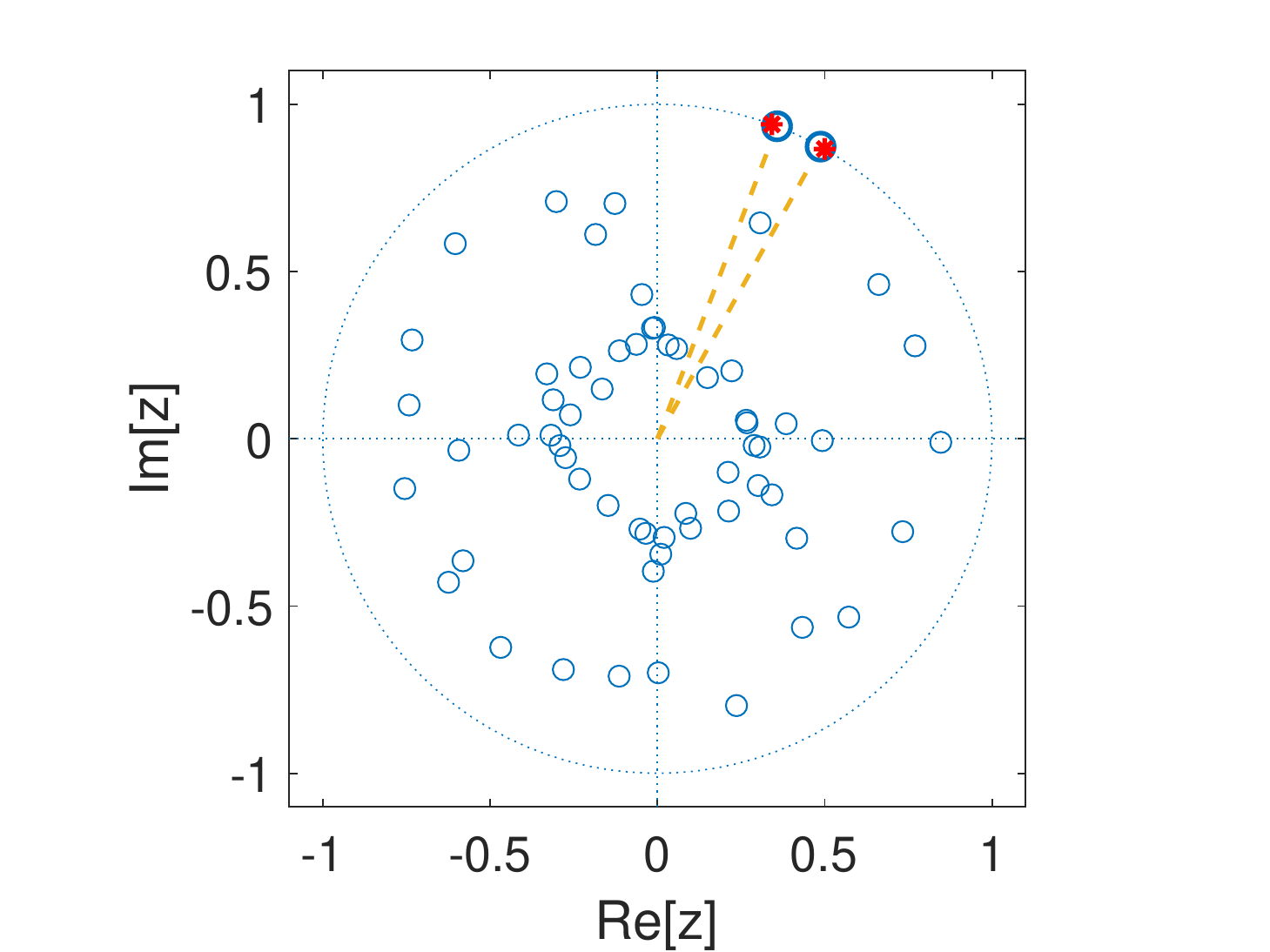} 
		\caption{Zeros of $p(z)$}
		\label{fig:Fig4b}
	\end{subfigure}
	\caption{Result for RPA with $ M=30$, $P=63 $, and two equal magnitude  sources with DOAs at $ 60^\circ$ and $70^\circ $ of $ 20 $\,dB SNR. }
	\label{fig:Fig4}
\end{figure}

In \cref{fig:Fig4a}, we consider an RPA with $ 30 $ sensors.
The minimum sensor spacing is $ d = \lambda/4 $, and the distance of the farthest sensor from origin is around $ 2\lambda $.
The proposed method resolves both sources as shown in \cref{fig:Fig4b}, 
whereas, the CBF results in a single peak at $ 65 ^\circ $ (CBF result not shown).

\vspace*{-5mm}

\begin{figure}[htbp]
	\begin{subfigure}{.48\columnwidth}
		\centering
		\includegraphics[width=1\linewidth]{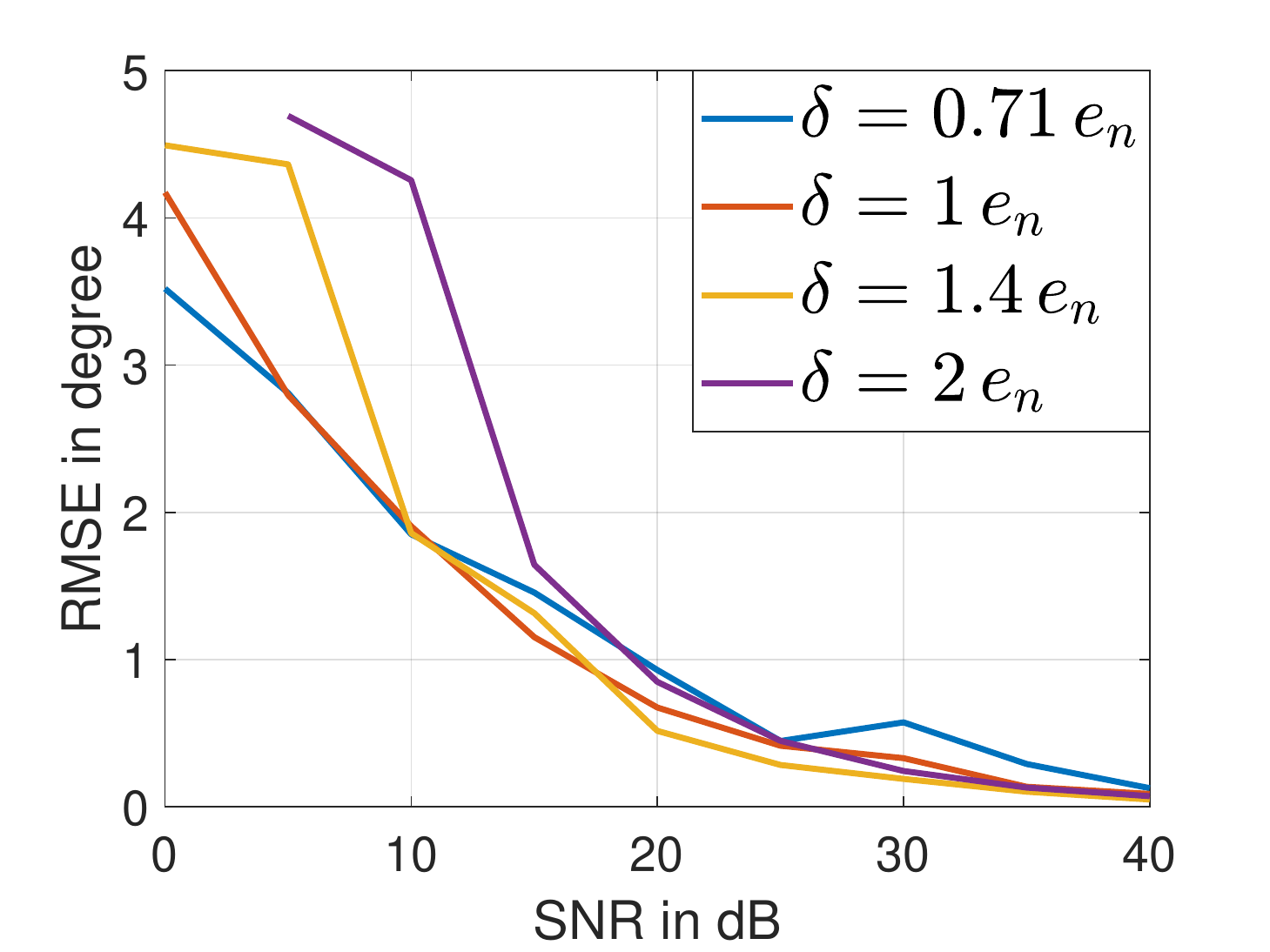}
		\caption{source separation = $ 10^\circ $ }
		\label{fig:Fig5a}
	\end{subfigure}
	\begin{subfigure}{.48\columnwidth}
		\centering
		\includegraphics[width=1\linewidth]{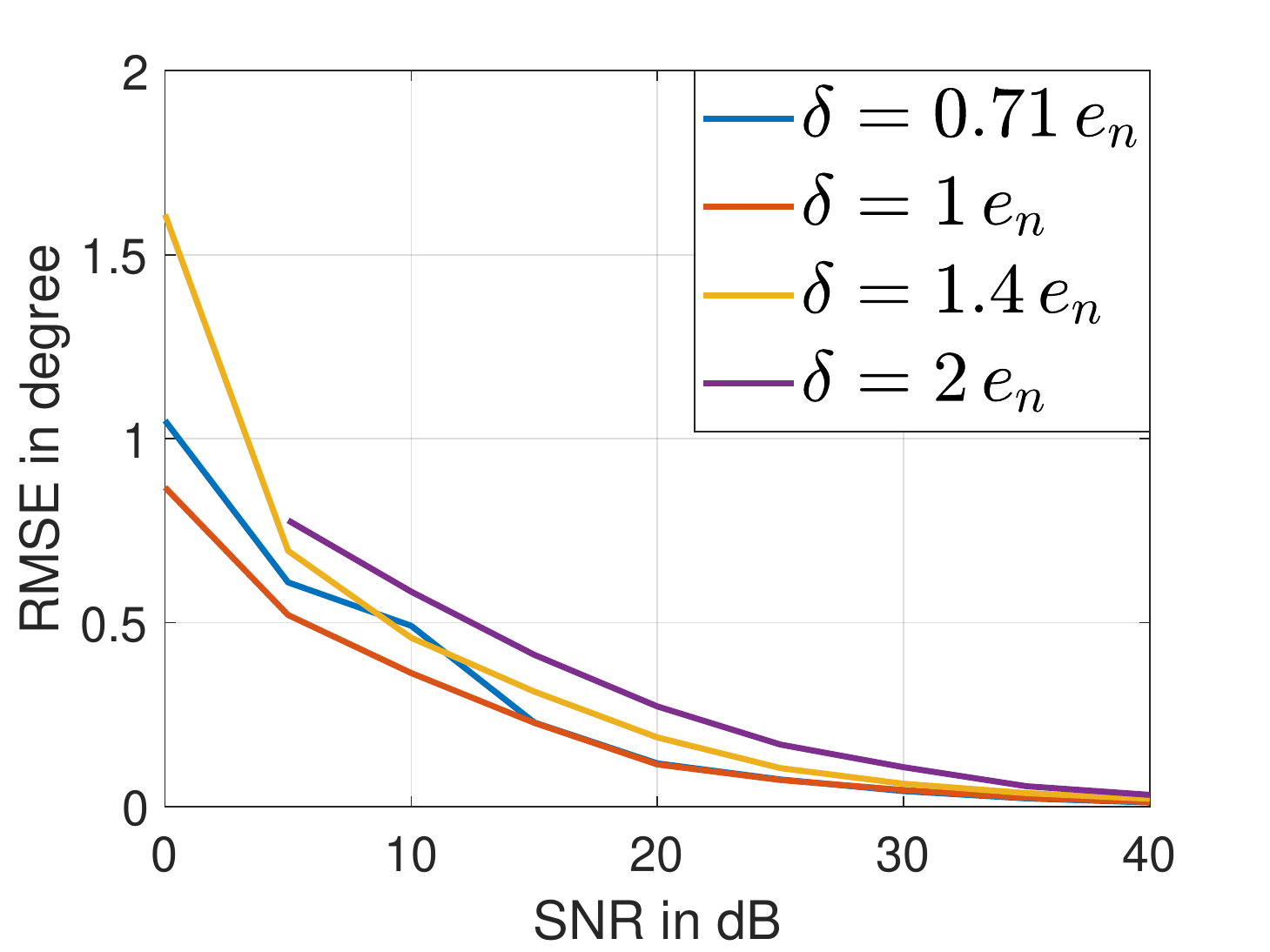}
		\caption{source separation= $ 30^\circ $ }
		\label{fig:Fig5b}
	\end{subfigure}
	\caption{DOA accuracy vs. SNR for UCA with $M=30$, $P=63 $. 50 trials, two sources of random DOAs in each trial.
	}
	\label{fig:Fig5}
\end{figure}
\subsection{Performance Evaluation Vs. SNRs }    
\vspace*{-1mm}
We now evaluate the performance of the approach for various SNRs,  and the sensitivity of the method to the value of noise norm upper-bound  $ \delta $.
The RMSE in DOA estimation for different SNRs and $ \delta $ is provided in \cref{fig:Fig5a,fig:Fig5b} for two sources of separation $ 10^\circ$ and $30^\circ $, respectively.
$ e_n $ is the expected value of noise norm $ e_n = \mathbb{E}[\lVert \bm{n}\rVert^{\phantom{.}}_2]$.
For i.i.d noise $ \mathcal{N} (0, \sigma_n ) $ across the sensors, $e_n = \sigma_n \sqrt{M} $.
The simulation considers 50 Monte Carlo trials with random source DOAs for each SNR. 
The performance depends on the minimum separation between sources. 
For larger separations, a smaller DOA error was observed. 
Note that at $ 10^\circ $ separation, the CBF is unable to resolve the sources at all SNRs (see \cref{fig:Fig2}).
Regarding the choice of $ \delta $, an underestimation of $ \delta $ was observed to cause many extraneous unit circle roots, but the $ \ell_1 $ recovery could remove those additional roots. 
The overestimation of $ \delta $, on the other hand, resulted in fewer roots on the unit circle, but they were slightly less accurate.
In general, with $ \ell_1 $ recovery processing, an underestimated $ \delta $ provided better results than the overestimated one.
As the SNR improves, the performance becomes less dependent on the choice of $ \delta. $
For SNR above $ 30 $\,dB in \cref{fig:Fig5b}, the estimates are nearly perfect.
The parameters involved in the approach are: $ \delta $, $ \beta $ and two thresholds, one for unit-circle zero detection, another for discarding low magnitude coefficients in the $ \ell_1 $ recovery.
\vspace*{-3mm}

\section{Discussion}
\vspace*{-2.3mm}
We have presented a search-free super-resolution DOA estimation and beamforming method for arbitrary geometry arrays, which is applicable for a single noisy snapshot, and correlated or uncorrelated sources. 
Further SNR improvement should be possible using multiple snapshots.
The upper bound of the noise norm $ \delta $ in \eqref{eq:gridfree_primal} needs to be estimated in practice. 
However, unlike traditional high resolution approaches, the proposed method does not require knowledge of the number of sources.
We made comparisons with the CBF, but not with traditional high resolution DOA approaches such as MUSIC and MVDR as they fail in the single snapshot case and coherent signal conditions, though they are applicable for arbitrary arrays. 
Moreover, existing sparsity based gridless super-resolution approaches are applicable only for ULAs. 
Simulation results prove that the new method can perform high resolution search-free DOA estimation for arbitrary geometries, 
using a single noisy snapshot.

\FloatBarrier
\bibliographystyle{IEEEtran} 
\ninept
\bibliography{Refs_forConf}
\end{document}